\documentclass[aps,prl,twocolumn,superscriptaddress,amsfont,graphicx,nofootinbib,preprintnumbers]{revtex4}
\usepackage{color,graphicx,epsfig}
\usepackage{ifpdf}
\usepackage{amsmath}
\usepackage{bm}
\usepackage{color}
\usepackage[english]{babel}
\usepackage{graphicx}
\usepackage{amsfonts}
\usepackage{amssymb}
\usepackage[bookmarks=false]{hyperref}
\usepackage{enumerate}

\definecolor{nicered}{rgb}{0.7,0.1,0.1}
\definecolor{nicegreen}{rgb}{0.1,0.5,0.1}
\hypersetup{colorlinks,citecolor= nicegreen,linkcolor= nicered}

\newcommand{\beq}{\begin{equation}}
\newcommand{\eeq}{\end{equation}}
\newcommand{\bea}{\begin{eqnarray}}
\newcommand{\eea}{\end{eqnarray}}

\definecolor{Red}{rgb}{1.,0.,0.}

\def\gsim{{~\raise.15em\hbox{$>$}\kern-.85em
          \lower.35em\hbox{$\sim$}~}}
\def\lsim{{~\raise.15em\hbox{$<$}\kern-.85em
          \lower.35em\hbox{$\sim$}~}}

\def\mysection#1{{{\bf #1}.~}}
\def\OMIT#1{}

\begin{document}

\def\Cincy{Department of Physics, University of Cincinnati, Cincinnati, Ohio 45221,USA}
\def\Perimeter{Perimeter Institute for Theoretical Physics, 31 Caroline St. N, Waterloo, Ontario, Canada N2L 2Y5}
\def\INFNRome{INFN, Sezione di Roma, Piazzale A. Moro 2, I-00185 Roma, Italy}


\title{Uncovering Mass Generation Through Higgs Flavor Violation}

\author{Wolfgang Altmannshofer}
\email[Electronic address:]{waltmannshofer@perimeterinstitute.ca}
\affiliation{\Perimeter}

\author{Stefania  Gori}
\email[Electronic address:]{sgori@perimeterinstitute.ca}
\affiliation{\Perimeter}

\author{Alexander L. Kagan}
\email[Electronic address:]{kaganal@ucmail.uc.edu}
\affiliation{\Cincy}

\author{Luca Silvestrini}
\email[Electronic address:]{Luca.Silvestrini@roma1.infn.it}
\affiliation{\INFNRome}

\author{Jure Zupan} 
\email[Electronic address:]{zupanje@ucmail.uc.edu} 
\affiliation{\Cincy}

\begin{abstract}
A discovery of the flavor violating decay $h\to \tau\mu$ at the LHC would require extra sources of electroweak symmetry breaking (EWSB) beyond the Higgs in order to reconcile it with the bounds from $\tau \to \mu\gamma$, barring fine-tuned cancellations. In fact, an $h \to \tau \mu$ decay rate at a level indicated by the CMS measurement is easily realized if the muon and electron masses are due to a new source of EWSB, while the tau mass is due to the Higgs. We illustrate this with two examples:  a two Higgs doublet model, and a model in which the Higgs is partially composite, with EWSB triggered by a technicolor sector.
The 1st and 2nd generation quark masses and CKM mixing can also be assigned to the new EWSB source.  Large deviations in the flavor diagonal lepton and quark Higgs Yukawa couplings are generic.
If $m_\mu$ is due to a rank 1 mass matrix contribution, a novel Yukawa coupling sum rule holds, providing a precision test of our framework.  Flavor violating quark and lepton (pseudo)scalar couplings combine to yield a sizable $B_s \to \tau \mu$ decay rate, which could be ${\mathcal O}(100)$ times larger than the SM $B_s \to \mu\mu$ decay rate.
\end{abstract}

\maketitle

Measurements of Higgs production and decay~\cite{:2012gk,:2012gu} have revealed that most of the electroweak symmetry breaking (EWSB) is due to the vacuum expectation value (vev) of the Higgs field. In the Standard Model (SM) the Higgs vev also sources the charged fermion masses. Testing this assumption directly is possible for the third generation fermions by measuring the Higgs decays to $b-$quarks and tau leptons, and by measuring the $t\bar th$ cross section at the LHC. Present measurements indicate that the Higgs is at least partially responsible for the masses of the $3^{\rm rd}$ generation fermions.
Much less is known about the origin of mass for the first two generations.  
There is experimental confirmation that the Higgs has a smaller Yukawa coupling to the muon than to the tau~\cite{Khachatryan:2014aep,Aad:2014xva}, as expected in the SM. The SM also predicts that the Higgs should not have tree level flavor changing decays, e.g., $h\to b s$ or $h\to \tau \mu$. The discovery of such decays would mean that there must be new physics (NP) near the electroweak scale~\cite{Pilaftsis:1992st,Korner:1992zk,Harnik:2012pb,Davidson:2010xv,Blankenburg:2012ex,Goudelis:2011un,Dorsner:2015mja,Crivellin:2015lwa,Crivellin:2015mga,Dery:2013rta,Kopp:2014rva,Dery:2014kxa,Sierra:2014nqa,Campos:2014zaa,Varzielas:2015iva,Lee:2014rba,He:2015rqa}.  In this Letter we point out that flavor violating Higgs decays can 
also be understood as a test of fermion mass generation, and we devise a sum rule that can be checked experimentally.

Intriguingly, the CMS collaboration has obtained the first bounds on Br$(h\to \tau \mu)<1.51\%$ at 95\% C.L., with a hint of a nonzero signal~\cite{Khachatryan:2015kon}. The best fit branching fraction is Br$(h\to \tau \mu)=(0.84_{-0.37}^{+0.39})\%$. 
We will show that the strength of this signal is naturally understood if a second source of EWSB is responsible for the muon mass.  This means that there is a whole family of NP models that can lead to large flavor violating Higgs decays.
We also extend this possibility to the quark sector.

\begin{figure}[t]
\centering{
\raisebox{1cm}{ a)}\hspace{-2mm}
\raisebox{-0.3mm}{ 
\includegraphics[scale=0.68]{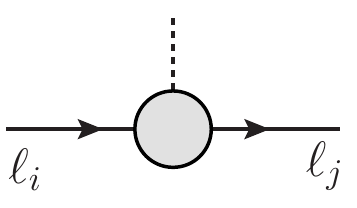}
}
\hspace{-3mm}
\includegraphics[scale=0.68]{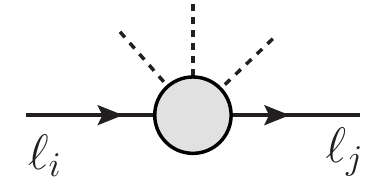} 
\raisebox{1cm}{ b)}\hspace{-3mm}
\raisebox{-4mm}{
\includegraphics[scale=0.68]{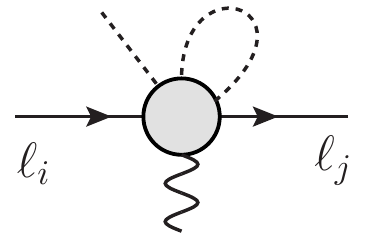}
}
\vspace{-4mm}
}
\caption{Contributions to the lepton mass matrix and Yukawa interactions (a) and the electromagnetic dipole (b).
\label{fig:single:Higgs}}
\end{figure}

\begin{figure}
\centering{
\raisebox{1cm}{ a)}\hspace{-2mm}
\raisebox{-0.3mm}{ 
\includegraphics[scale=0.54]{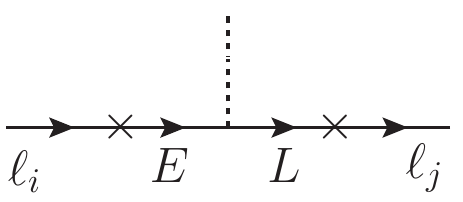}
}
\hspace{-3mm}
\includegraphics[scale=0.54]{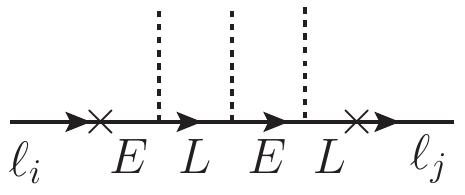} 
\raisebox{1cm}{ b)}\hspace{-3mm}
\raisebox{-2mm}{
\includegraphics[scale=0.54]{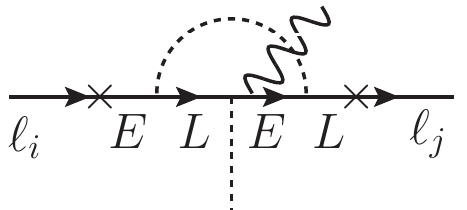}
}
\vspace{-4mm}
}
\caption{A realization of Fig. \ref{fig:single:Higgs} with vectorlike leptons. 
\label{fig:single:Higgs:vectorlike}}
\end{figure}

Let us first discuss $h \to \tau \mu$ in models in which the Higgs is the only source of EWSB.
In an effective field theory, in which the NP particles are integrated out, the Higgs-lepton couplings are \cite{Goudelis:2011un,Dorsner:2015mja}
\beq\label{eq.Higgsoperator}
- {\cal L}_{\rm Y}= \lambda_{ij}(\bar \ell_L^i e_R^j) H + \frac{\lambda_{ij}'}{\Lambda^2} (\bar \ell_L^i e_R^j) H (H^\dagger H)+{\rm h.c.},
\eeq 
where $\Lambda$ is the NP scale, and we have kept the two leading terms. 
In Fig.~\ref{fig:single:Higgs} a) the two operators are denoted with a blob corresponding to the exchange of NP states.
For example, the latter could be vectorlike leptons of mass $\Lambda$ which mix with the SM leptons, 
see Fig. \ref{fig:single:Higgs:vectorlike} a) (Note that if the only NP states are scalars, then \eqref{eq.Higgsoperator} implies the presence of additional EWSB vevs \cite{Heeck:2014qea}.).

A misalignment of $\lambda_{ij}$ and $\lambda_{ij}'$ in flavor space leads to off-diagonal Higgs Yukawa couplings in the mass basis. Using the normalization in \cite{Harnik:2012pb}, we find
\beq
Y_{\tau \mu  }= \frac{v_W^2}{\sqrt{2} \Lambda^2}\langle \tau_L  | \lambda'  | \mu_R \rangle, \label{eq:Ytaumu}
\eeq
and similarly for $Y_{\mu\tau}$, with the Higgs vev $v_W = 246$ GeV. 
The CMS measurement \cite{Khachatryan:2015kon} gives
\beq 
\sqrt{|Y_{\tau \mu}|^2+|Y_{\mu\tau}|^2}=(2.6\pm 0.6 )\cdot 10^{-3}\,. \label{CMSrange}
\eeq 
 
In the blobs of Fig.~\ref{fig:single:Higgs} at least one NP particle needs to carry 
electromagnetic charge.  Thus, the electromagnetic dipole operators,
\beq\label{eq:dipole:ops}
{\cal L}_{\rm eff}=c_{L,R} \, m_\tau \frac{e }{8\pi^2} \big(\bar \mu_{R,L} \sigma^{\mu\nu}\tau_{L,R}\big) F_{\mu\nu},
\eeq
are also generated via photon emission from intermediate NP states.
Estimating the amplitude in Fig.~\ref{fig:single:Higgs}~b) using na\"{i}ve dimensional analysis (NDA)
gives 
\beq\label{eq:NDA:cLR}
c_{L}\sim \frac{v_W}{\sqrt 2 m_\tau \Lambda^2} \langle \tau_L  | \lambda'  | \mu_R \rangle = \frac{Y_{\tau\mu}}{m_\tau v_W},
\eeq
and similarly for $c_R$.
The bound Br$(\tau \to \mu\gamma)<4.4\cdot 10^{-8}$ (90\% C.L.)~\cite{Aubert:2009ag} implies
\beq
\sqrt{|c_L|^2+|c_R|^2}< (3.8~{\rm TeV})^{-2} .
\eeq
Comparing with \eqref{eq:NDA:cLR}, and taking $Y_{\tau \mu}\sim Y_{\mu\tau}$, yields 
\beq
\sqrt{|c_L|^2+|c_R|^2}\sim \left(\frac{Y_{\tau\mu}}{2.2\cdot 10^{-5}}\right)(3.8~{\rm TeV})^{-2},
\eeq
which generically excludes the observed $h\to \tau \mu$ rate
by four orders of magnitude (see \eqref{CMSrange}), as observed in the vectorlike lepton case~\cite{Falkowski:2013jya,Dorsner:2015mja}.  
We conclude that the observed $h\to \tau \mu$ rate can only be explained if either (i) 
$\tau \to \mu \gamma$ is suppressed by apparently fine-tuned cancellations, or (ii) the Higgs is not the only source of EWSB.  

Specifically, we will show that the observed $h\to \tau \mu$ rate can be explained
in models in which the lepton mass matrix is of the form
\beq{\cal M}^\ell={\cal M}_0^{\ell}+\Delta {\cal M}^\ell, \label{eq:M:decomp}
\eeq
where a rank 1 matrix ${\cal M}^\ell_0$ is due to the vev of a scalar $\phi$ (the primary component of the Higgs), and accounts for the bulk of $m_\tau$. The matrix $\Delta {\cal M}_\ell$ is due to an additional source of EWSB, can be rank 2 or 3, and accounts for $m_{e}$ and $m_\mu$.  We first focus on the 2nd and 3rd generations. We choose the flavor basis in which  $({\cal M}_0^{\ell})_{33} \sim m_\tau$ is the only non-zero entry of ${\cal M}^\ell_0$, so that generically
 \beq (\Delta {\cal M}^\ell)_{ij} = {\cal O}(m_\mu ),~~ i,j=2,3. \label{naturalmasses}\eeq
The flavor violating Yukawa couplings are given by
\beq  
v_W Y_{ \mu\tau}  = -R_{ Y} \,(\Delta {\cal M}^\ell )_{\mu\tau} \,,\label{YFCleptons}
\eeq
and similarly for $Y_{ \tau\mu }$. Here $(\Delta {\cal M}^\ell )_{\mu\tau} \equiv \langle \mu_L | \Delta {\cal M}^\ell |\tau_R  \rangle$, while $R_Y$ only depends on the details of the EWSB sector. 
Taking $(\Delta {\cal M}^\ell )_{\mu\tau} \sim (\Delta {\cal M}^\ell )_{\tau\mu} $ and $R_Y \sim 1$, the $h \to \tau \mu$ rate \eqref{CMSrange} corresponds to 
$ (\Delta {\cal M}^\ell )_{\mu\tau}  \sim  (0.45 \pm 0.10)~{\rm GeV}$,
consistent with  \eqref{naturalmasses}. 

If there is more than one contribution to ${\cal M}^l$ the $\tau \to \mu\gamma$ constraint is easily satisfied. For instance, if $\Delta {\cal M}^\ell$ originates from a radiative or new strong interaction form factor at a NP scale $\Lambda$
the dipole operator coefficients \eqref{eq:dipole:ops} generically scale as 
\beq
c_{L,R} \sim \frac{(\Delta {\cal M}^\ell )_{\mu\tau,\tau\mu} }{\Lambda^2} \frac{8 \pi^2 }{m_\tau } \sim  \frac{Y_{\tau \mu,\mu\tau}}{ m_\tau v_{W}}\frac{8 \pi^2 v_{W}^2}{\Lambda^2}. \label{eq:cLRscaling}
\eeq
Compared to \eqref{eq:NDA:cLR}, there is an extra factor ${8 \pi^2 v_{W}^2}/{\Lambda^2}$. Thus,
consistency with $\tau \to \mu \gamma$ can always be achieved for sufficiently large  $\Lambda \ge {\mathcal O}(10)$ TeV. 

We also consider the analog of \eqref{eq:M:decomp}-\eqref{YFCleptons} for quarks with the same two sources of EWSB and, therefore, with the same $R_Y$. 
It is natural to consider
$\Delta {\cal M}^{u,d}_{ij} = {\mathcal O}(m_{c,s} )$ 
for $i,j=2,3$. 
Generation of $m_c$, $m_s$ and $V_{cb} $
then implies  
\beq
\begin{split} 
(\Delta {\cal M}^{u,d})_{22} \approx m_{c,s},
~~~(\Delta {\cal M}^d)_{23}  \approx V_{cb} \,m_b,~~~~~~~
\end{split} \label{msVcb}
\eeq
and $R_Y^2 \Delta {\cal M}^d_{32}   \lesssim V_{cb} \,m_b / 6$ from the bound on the (Higgs exchange) $B_s$ mixing operator $(\bar b_R s_L ) (\bar b_L s_R)$ \cite{Bevan:2014cya}.

An example of a model that can produce the structure in \eqref{eq:M:decomp} and the corresponding one in the quark sector is a two Higgs doublet model (2HDM). (In previous 2HDM studies of the $h \to \tau \mu$ signal, $m_\mu$ was due to the Higgs vev~\cite{Crivellin:2015lwa,Crivellin:2015mga,Campos:2014zaa,Heeck:2014qea,Sierra:2014nqa}.)   
The Higgs doublets $\Phi$ and $\Phi'$ contain the neutral scalars $\phi$ and $\phi'$, with vevs $v$ and 
$v'$, respectively, where $v_W^2 = v^2 + v'^2 $.  
The field $\phi$ has a Yukawa coupling $\phi\, \bar \ell^3_L e^3_R $, whereas $\phi'$ has couplings 
to all three families, consistent with \eqref{naturalmasses}.
Note that a hierarchy in the vevs, $v \gg v' $, can help explain the mass ratio $m_\mu / m_\tau $.   
The Yukawa coupling structure can, for instance, follow from a symmetry
which is horizontal, or which distinguishes between new vector like leptons and the SM ones~\cite{LongPaper,Bauer:2015fxa}.  The two Higgs doublets would transform differently, equivalent to a Peccei-Quinn (PQ) symmetry that is softly broken by the $m^2 \phi \phi'$ term,   as required by vacuum alignment.

The off-diagonal Higgs Yukawa couplings satisfy~\eqref{YFCleptons}, with
$R_Y$ given by
\beq R_Y = R_{\alpha\beta} \equiv 2 \cos(\alpha-\beta) \big/ \sin 2\beta \,. \label{Ralphabeta}\eeq
Here, the ratio of vevs is defined as $\tan\beta = v /v'$, and the mixing of $\phi$ and $\phi'$ yields the light and heavy Higgs mass eigenstates $h= \phi \cos\alpha- \phi' \sin\alpha$, $H=\phi \sin\alpha +\phi'\cos\alpha $.
The reduced flavor diagonal Yukawa couplings $\hat y_{a} \equiv  Y_{aa}/ Y_{aa}^{\rm SM} $ are given by
\beq  \hat y_a  = \cos\alpha/\sin\beta - R_Y (\Delta {\cal M}^\ell )_{aa}/m_a ,~~a=\mu,\,\tau, \label{flavdiag}\eeq
valid in the phase convention $m_{a} \equiv ({\mathcal M}^\ell)_{aa} >0$. 

The 2HDM with tree level Yukawa couplings provides an exception to the scaling in \eqref{eq:cLRscaling}.  
It satisfies the bound from $\tau\to\mu\gamma$ due to an additional $y_\tau$ insertion compared to \eqref{eq:NDA:cLR} and heavy Higgs mass supression \cite{Davidson:2010xv}.  
Variations in which the $\phi'$ Yukawa couplings are radiatively induced would possess the scaling in \eqref{eq:cLRscaling}.

Horizontal symmetries may imply that certain $\phi'$ Yukawa couplings vanish.  For example, the charges of a global $U(1)$ symmetry,
or a simple $Z_3$ in the two generation case, can be chosen such that
$\Delta {\cal M}^\ell$ only has off-diagonal nonzero entries, $m'_{23}$ and $m'_{32}$.
We refer to this example as the ``horizontal'' case.  We also consider a ``generic'' case, in which 
all $m'_{ij} $ can be non-zero.

\begin{figure}[t]
\centering{
\includegraphics[scale=0.42]{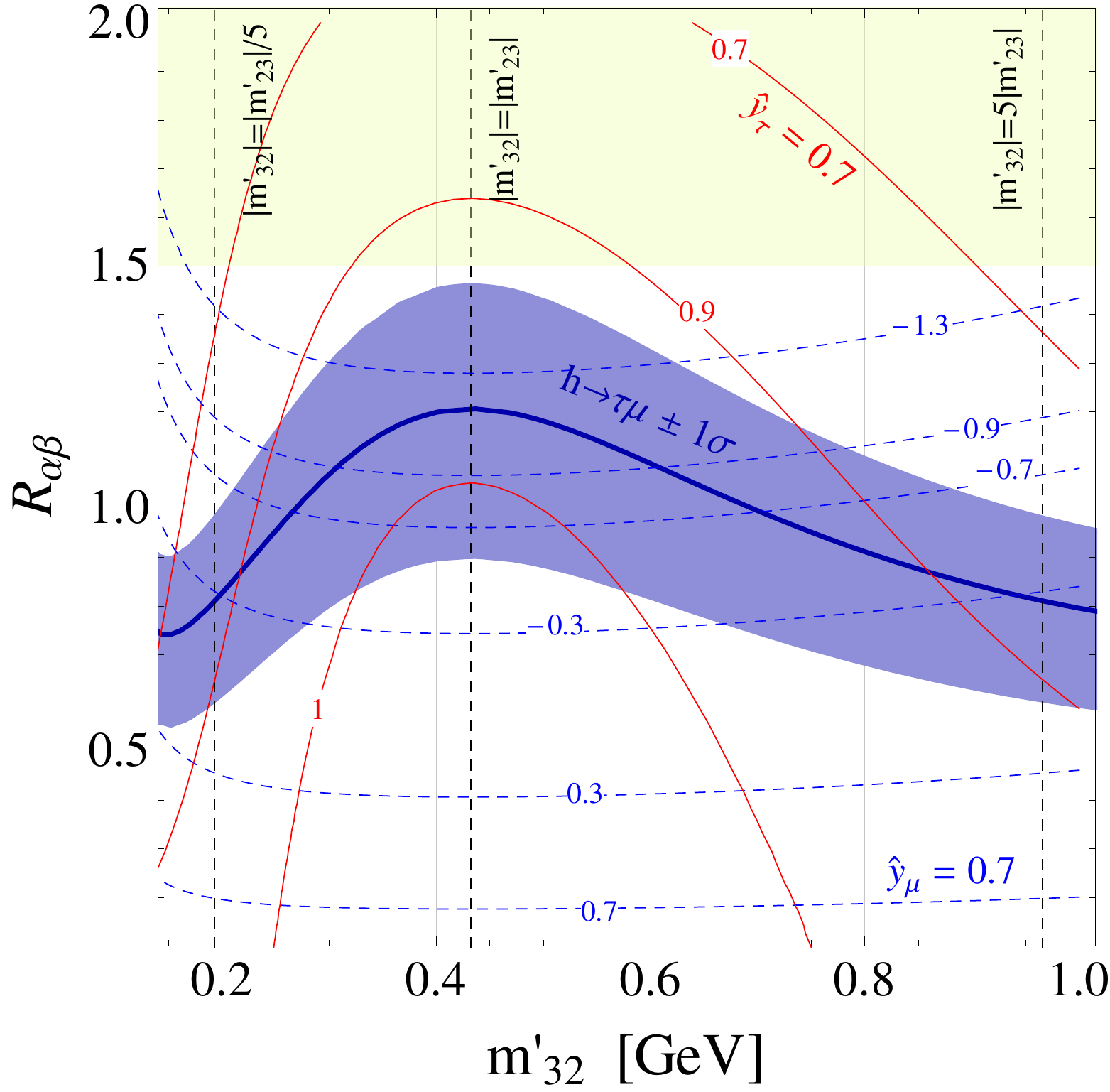}
}
\caption{The region favored by the measurement of Br($h\to\tau\mu$) at the $1\sigma$ level (in blue). The dashed vertical lines correspond to $|m_{32}'/m_{23}'|=1/5,1,5$.
Contours of $\hat y_\mu $ (blue) and $\hat y_\tau $ (red) are shown for $\tan\beta = 2$. The yellow region is in conflict with the measurement of the $hZZ$ coupling, for $\tan\beta = 2$.
\label{fig:Ralphabeta}}
\end{figure}
 
In the horizontal case, two of the entries in ${\cal M}^\ell $ are fixed by $m_{\mu}$ and $m_{\tau}$, leaving one free parameter, taken to be $m_{32}'$.
The Higgs couplings are fixed by $m_{32}'$ and the angles $\alpha$,$\,\beta$. 
Fig.~\ref{fig:Ralphabeta} shows the region in the $m'_{32}$ - $R_{\alpha\beta}$ plane favored by the CMS result in \eqref{CMSrange}. (A similar range of 
$R_{\alpha\beta}$ is spanned in the generic case).
The Higgs coupling to weak gauge bosons ($g_{hVV}$) is modified by a factor $\sin(\beta-\alpha)$.
For $R_{\alpha\beta} > 1.5$ and $\tan\beta=2$, the shift satisfies $|\delta g_{hVV} / g^{\rm SM}_{hVV} |\gsim 20\%$, in conflict with Higgs data. For larger $\tan\beta$, this constraint on $R_{\alpha\beta}$ is weakened.

From Fig.~\ref{fig:Ralphabeta}, the CMS result requires $R_{\alpha\beta}= {\mathcal O}(1)$, versus the decoupling limit $R_{\alpha\beta} \to 0$.
Expanding in $v_W^2/m_A^2$ and $1/\tan\beta$, with $A$ the neutral pseudoscalar, 
\begin{equation} R_{\alpha\beta} \simeq v_W^2/m_A^2\times(\lambda_3+\lambda_4  + ... )  \, \label{lambda34} \end{equation}
in the PQ symmetric limit $\lambda_ {5,6,7}=0$ (we use the notation of \cite{Haber:1989xc} for the quartic scalar couplings, $\lambda_i $). 
The value $R_{\alpha\beta} \sim 1$ can be obtained with $m_A \sim 500$\,GeV and $\lambda_3 \sim \lambda_4 \sim 2$. 
Such couplings are compatible with electroweak precision constraints, and do not develop Landau poles below $\mathcal O( 30)$ TeV.  For smaller $\lambda_{3,4}$ the poles can be pushed  beyond $M_{\rm GUT}$ while maintaining $R_{\alpha\beta} \sim 1$, if $\lambda_7 \ne 0$ due to PQ symmetry breaking.  In that case, at large $\tan\beta$, $\Delta R_{\alpha\beta}\sim v_W^2/m_A^2\times (\lambda_7\tan\beta)$, which 
could originate, e.g., from a dimension 5 coupling $|\phi |^2 \phi \phi' S$ to a PQ charged singlet scalar $S$, as in the NMSSM.

\begin{figure}[t]
\includegraphics[scale=0.37]{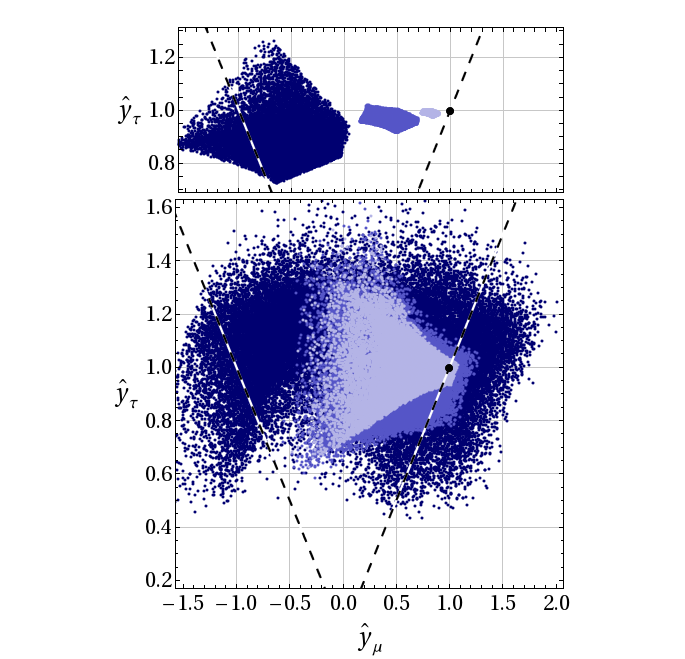}
  \caption{The reduced Higgs couplings $\hat y_\mu$ and $\hat y_\tau$ for the horizontal case (top panel), generic case (bottom panel), and SM (black dot). Dark blue, blue and light blue regions reproduce the CMS Br($h\to\tau\mu$) measurement, 1/3 of it and 1/10 of it, at the $1\sigma$ level.  The dashed lines satisfy $\hat y_\mu /\hat y_\tau  =\pm 1$.} \label{fig:scatterPlots}
\end{figure}

Observable $h \to \tau \mu$ is correlated with significant deviations of the flavor diagonal couplings from their SM values, as can already be seen in Fig.~\ref{fig:Ralphabeta}.
Fig.~\ref{fig:scatterPlots} shows $\hat y_{\mu} $ vs. $\hat y_\tau$
for ``horizontal'' 
and ``generic'' parameter scans. We take $1/5 < |m'_{32}/m'_{23}| < 5$ in the horizontal case (corresponding to $0.2$\,GeV $\lesssim m_{32}' \lesssim 0.95$\,GeV in Fig.~\ref{fig:Ralphabeta}), and $|(\Delta {\cal M}^\ell)_{ij} | < 5m_\mu$ for all entries in the generic case.
Both scans allow $\lambda_{3,4}\le 2$, $m_A \ge 400$\,GeV, $|\delta g_{hVV} /g_{hVV}^{\rm SM}| \le 20\%$, and a heavy Higgs production cross section below $ 10\% $ of a SM Higgs with same mass, to be consistent with heavy scalar direct search bounds.

In the horizontal case, the CMS result implies a negative $\hat y_\mu$, with $|\hat y_\mu |$ typically well below 1, and $|\hat y_\tau -1 |\lsim  25\%$.
The deviations tend to be larger in the generic case.
The ratios $|\hat y_\mu | < 1 $ and $| \hat y_\mu / \hat y_\tau | < 1$ (vs. $\hat y_\mu / \hat y_\tau \approx 1$ in the type-II 2HDM) are favored in the current, as well as hypothetical future scenarios with a $3\times$ or $10\times$ smaller $h\to\tau\mu$ rate (and scaled $1\sigma$ errors).

If the quark Yukawa coupling structure in the 2HDM is analogous to \eqref{eq:M:decomp}, 
with $v'$ yielding \eqref{msVcb}, then the off-diagonal quark couplings satisfy $Y_{ct,tc} = {\mathcal O}(m_c /v_W)$, $Y_{bs}\ll Y_{sb}  \approx   5 \cdot 10^{-4} R_Y$, see \eqref{msVcb} and below.   
There are new contributions to $B_s \to \mu \mu $, with $A$ exchange being the largest \cite{LongPaper}.
In the horizontal case, the ${\rm Br}(B_s \to \mu \mu )$ measurement~\cite{CMS:2014xfa} constrains $\tan\beta$, e.g. $\tan\beta \lesssim 7$ for $m_A \simeq 500$\,GeV. In the generic case much larger values of $\tan\beta $ 
are allowed.
The $B_s \to \mu\mu$ bound has been imposed in Fig.~\ref{fig:scatterPlots}. 
Roughly 80\% of the points do not require tuned cancelations in
$m_\mu $ and $B_s \to \mu\mu$. 
From \eqref{flavdiag}, the diagonal couplings satisfy
$\hat y_{c,s} = \cos\alpha/\sin\beta - R_Y$ and $\hat y_{t,b} = \cos\alpha/\sin\beta$,
up to ${\cal O}(m_{c,s}/m_{t,b} )$.
Thus, while $\hat y_{t,b}$ receive modest corrections $\le  20\% $, $\hat y_{c,s}$ tend to be ${\cal O}(1)$ suppressed, and could even vanish in tuned regions of parameter space.
This possiblity, given a new source of light quark masses, has been mentioned in~\cite{Perez:2015aoa}.

In our next illustration of~\eqref{eq:M:decomp},
$\Delta {\cal M}_\ell$ is due to technicolor (TC) strong dynamics.  
The Higgs is a mixture of $\phi$ and a composite heavy scalar, $\sigma_{\rm TC}$.  As in the 2HDM, in addition to the heavy Higgs state ($H$) there is a charged 
scalar and a neutral pseudoscalar ($A$) (both also partially composite).
The framework is bosonic technicolor (BTC)~\cite{Simmons:1988fu,Samuel:1990dq,Dine:1990jd,Carone:1992rh,Carone:1993xc,kagantalk,Antola:2009wq,Azatov:2011ht,Chang:2014ida}, motivated by improved naturalness of EWSB in supersymmetric models.  For simplicity, we consider the non-supersymmetric case. We add to the SM a weak doublet and two weak singlet technifermions, $T_R=(U_R , D_R )^T $ and $D_L, U_L$, and a
technicolored scalar~\cite{Kagan:1991ng,Kagan:1994qg,Dobrescu:1995gz}, $\xi$, all transforming in the fundamental of the confining TC  gauge group, e.g. $SU(2)_{\rm TC}$.
%
TC confinement yields the $SU(2)_L$ breaking condensates $\langle \bar DD \rangle, \,\, \langle \bar UU \rangle$ at a scale $\Lambda_{\rm TC}\sim 4\pi f_{\rm TC}$, where $f_{\rm TC}$ is the technipion decay constant. 
The $W$ and $Z$ masses receive contributions from TC and the Higgs, so that $ v_W^2 \simeq f_{\rm TC}^2 + v^2 $, where $\langle \phi \rangle = v$ is a Higgs vev.
The Higgs and precision electroweak phenomenology is viable if $f_{\rm TC}\lsim 80$ GeV \cite{Chang:2014ida,Kaganetal:2020},
or $\tan\beta \equiv v/f_{\rm TC} \gsim 3$.
 
The effective operators
\beq  
\frac{{{h}}^{\ell}_i {{{h}}^e_j}^{\dagger}}{m_\xi^2 } ~\bar \ell_{L}^i T_R \bar D_L  e_{R}^j  + {\rm h.c.}, \label{4fermiops}
\eeq
follow from integrating out the $\xi$ field 
in the Yukawa couplings ${{h}}^{\ell}_i \xi    \bar  \ell^i_ L { T_ R }+ {{{h}}^e_i}^{\dagger}\xi ^* \bar D_L { e^i_{R }}$.
The TC condensates thus yield a rank 1 contribution to $\Delta {\cal M}^{\ell}$.
Employing a leading order chiral Lagrangian, we obtain the lepton masses and  
dipole coefficients \cite{LongPaper},
\beq
\!\!\!\! ({\Delta} {\cal M}^{\ell})_{ij}  =\eta\,  \kappa  \,
\frac{4 \pi f_{\rm TC}^3}{2 m_{\xi}^2} h^\ell_i \,h^{e\,\dagger}_j;
~~\frac{c_L}{8 \pi^2}=  Q_\xi \frac{(\Delta {\cal M}^{\ell})_{\tau\mu }}{2 m_\xi^2 m_\tau }
,\label{BTCmassanddipole}
\eeq
and similarly for $c_R$, where $Q_\xi=1/2$ is the $\xi$ electric charge, 
$\kappa  \sim 1.5$ based on  $1/N_c$ scaling from $n_f  \!=  \!2$ lattice QCD \cite{Baron:2009wt}, and $\eta$ accounts for RGE running between $\mu \sim m_\xi $ and $\mu \sim \Lambda_{\rm TC}$.
Given
the central value (less $1\sigma$) of the $h \to \tau \mu$ measurement,
consistency with the $\tau \to \mu \gamma $ bound requires $\sqrt{R_Y} \,m_\xi  \gsim 10\, (8.7)$ TeV.

The chiral Lagrangian yields $R_Y > \cos\alpha/\sin\beta $ to all orders in the chiral expansion \cite{LongPaper}, where $\alpha$ is the $\phi-\sigma_{\rm TC}$ mixing angle.  Given that $\cos\alpha \approx 1$ (due to the relatively large $\sigma_{\rm TC}$ mass) and $\sin\beta = v/v_W \approx 1$, $R_Y \!>\!1 $ to good approximation.  Using NDA, we obtain $R_Y -1 \sim 0.2 $, with large uncertainty due the poorly known mass and couplings of the $\sigma_{\rm TC}$.  

Numerical examples consistent with the $\tau \to \mu \gamma$ bound are easily found.
For instance, for ${ h}^\ell \!= \!{h}^e$, the CMS result (less $1\sigma$) is obtained for 
$h_3 \!\approx \!2.1(1.5)$ and $h_2 \!\approx\! 0.6(0.6)$ at the matching scale $\mu \sim m_\xi$. 
Alternatively, for $h_3^\ell = 0$, 
the signal (less $1\sigma$) is obtained if $h_{2}^\ell h_2^e \approx 0.6 (0.4)$ and $h_2^\ell h_{3}^e\approx 2.5 (1.5)$.
In these examples $R_Y \!= \!1.3$, $f_{\rm TC}\! = \!80$ GeV, $\eta \sim 3$ based on two loop estimates in $\alpha_{\rm TC}$, and $m_\xi \approx 8.8 \,(7.6)$ TeV, yielding 
${\rm Br}(\tau \to \mu \gamma)$ at the bound.  

The flavor diagonal couplings generically show large deviations from the SM predictions. In the above examples,
$\hat y_\mu $ is negative with magnitude ranging from  $\approx 0.2 - 0.9$, $\hat y_\tau \approx  0.9-1.6$,
and $|\hat y_\mu /\hat y_\tau | \approx (0.2 - 0.6)$, well below 
the SM and type-II 2HDM ratio. 

We extend \eqref{eq:M:decomp} to the 
quark sector via the colored techniscalar $\omega$
with couplings to the quark doublets ($h^q$) and quark singlets ($h^{u,d}$) analogous to 
$h^\ell $ and $h^e$, respectively \cite{LongPaper}.  Rank 1 $\Delta {\cal M}^{u,d} $ 
follow in analogy with \eqref{4fermiops}, \eqref{BTCmassanddipole}.  
Consistency with \eqref{msVcb} and with the bound on ${\rm Br}(b \to s \gamma )$ requires a scale $m_\omega \gsim 5$ TeV, 
similar to the $\tau \to \mu\gamma$ bounds on $m_\xi$. 
In turn, the quark masses and mixings can be obtained with all $h^{u,d}_i \lsim 1$.
The flavor diagonal Yukawa couplings satisfy
$\hat y_{c,s} \approx 1 -R_Y $ and $\hat y_{t,b} \approx 1 $, given $\cos\alpha/\sin\beta \approx 1$, see \eqref{flavdiag}.

Our general framework \eqref{eq:M:decomp} readily extends to three generations 
\cite{LongPaper}. 
For instance, 
in the flavor basis of \eqref{naturalmasses}, 
it is natural that $(\Delta {\cal M}^\ell )_{1i,i1} = {\mathcal O}(m_e )$.  
The couplings $Y_{e x , x e}$ ($x \!=\! \mu,\tau$)
then yield Higgs mediated 
$\mu \to e\gamma$ rates below the current bound. 
In the quark sector, with $(\Delta {\cal M}^{u, d)})_{1i,i1} = {\mathcal O}(m_{u,d} )$, 
consistency of the Higgs mediated FCNC's, e.g., $\,\epsilon_K$~\cite{Bevan:2014cya}, with $\theta_c$, $V_{ub}$ requires $(\Delta {\cal M}^{d})_{ji} \lsim (\Delta {\cal M}^{d})_{ij} /10$ ($[ij] =13,23$).  
These relations could result from horizontal
symmetries which address the fermion mass and mixing 
hierarchies.  It is noteworthy that $s\to dg$ dipole operators, with scaling analogous to \eqref{eq:cLRscaling}, 
could play a role in $\varepsilon^\prime/\varepsilon$, bridging the gap between experiment \cite{Batley:2002gn,AlaviHarati:2002ye,Worcester:2009qt} and the SM prediction \cite{Bai:2015nea,Buras:2015yba}.

A novel Yukawa coupling sum rule holds   
if $\Delta {\cal M}^\ell $, like ${\cal M}_0^\ell $, is rank 1 when neglecting the first generation.  This is the case in the BTC example, and could be realized more generally in 
the  
``rank 1'' approach to the fermion mass and mixing hierarchies, see e.g. \cite{Balakrishna:1987qd,Balakrishna:1988ks,Balakrishna:1988bn,Kagan:1989fp,Dobrescu:2008sz,Ibarra:2014fla,Baumgart:2014jya,Altmannshofer:2014qha}.
The sum rule is given by
\beq \hat y_\mu \hat y_\tau -  \hat y_{\tau \mu} \hat y_{\mu\tau}   = \hat y_{t , b} (  \hat y_\mu + \hat y_\tau - \hat y_{t  , b} )\,,\label{sumrule}\eeq
where $\hat y_{ij} \equiv Y_{ij} /Y_{ii}^{\rm SM} $, and we have substituted $\cos\alpha/\sin\beta \to \hat y_{t (b) }$, see \eqref{flavdiag}.
It holds up to corrections of ${\mathcal O}(m_{c } /m_{t },m_{s} /m_{b},m_{e} /m_{\mu})$.
Remarkably, the sum rule offers a precision test of
the rank 1 hypotheses, potentially validating
our framework in this case. If $\Delta {\cal M}^\ell $ has full rank, (\ref{sumrule}) holds up to ${\mathcal O}(m_{\mu } /m_{\tau })$ corrections, which can be sizable for large
$Y_{\tau\mu,\mu\tau}$ as in~(\ref{CMSrange})~\cite{LongPaper}.

Generation of the CMS $h \to \tau \mu$ result and $V_{cb} $ \eqref{msVcb} in our framework can lead to a sizable $B_s \to \tau \mu$ rate via $h$, $A$ and $H$ tree-level exchanges.  The $A$ and $H$ contributions grow as $(\tan\beta)^4$, whereas the $A$ contribution to ${\rm Br}(B_s \to \mu\mu )$ grows as $(\tan\beta)^2$ and tends to interfere destructively with the SM.   Thus, large values of the ratio $R_{\tau\mu} \equiv {\rm Br} (B_s \to \tau \mu)/{\rm Br} (B_s \to \mu \mu)_{\rm SM} $ are possible, without tuned cancelations in ${\rm Br}(B_s \to \mu\mu)$.
In our 2HDM and BTC examples, at moderate $\tan\beta \lsim 4$ and for $m_A, m_H \gsim 400$ GeV,
$R_{\tau\mu}  \lsim 10$ correlates with $\lsim 50\% $ suppression of Br$(B_s \to \mu\mu)$.  
However, for $\tan\beta \sim 6-10$, easily realized in the 2HDM, much larger $R_{\tau\mu}$ are possible:
in the generic (horizontal) case, $R_{\tau\mu}$ can be as large as $\sim 200$ ($\sim 50$)
accompanied by $\sim 50\%$ suppression ($\sim 20\%$ enhancement) of ${\rm Br}(B_s \to \mu\mu)$.
We estimate that Br$(B \to K^{(*)} \mu\tau)$ can be as large as $\mathcal{O}(10^{-7})$ in such cases.
The above framework could lead to potentially observable $t\to h c$ decays \cite{LongPaper} if, e.g.,
$V_{cb}$ receives a sizable contribution via
$(\Delta {\cal M}^u )_{23} = O(V_{cb} m_t ) $.

In summary, an observable $h \to \tau \mu$ signal is naturally realized in models where the 1st and 2nd generation masses and CKM mixing are due to a second source of 
EWSB. 
We have focused on the 2nd and 3rd generations, illustrating our framework with a two Higgs doublet model, and an example with a partially composite Higgs, where EWSB is triggered by new strong interactions.  
 The flavor diagonal Higgs Yukawa couplings typically show large deviations from the SM.  
 Finally, 
(pseudo)scalar exchanges can yield ${\rm Br}(B_s
 \to \tau \mu ) \lsim  {\rm few}  \cdot 10^{-7}$ and significant shifts in ${\rm Br}(B_s
 \to \mu \mu )$, both potentially detectable at the LHC.

\mysection{Acknowledgements}
The work of A.K. is supported by the DOE grant DE-SC0011784. J.Z. is supported in part by the U.S. National Science Foundation under CAREER Grant PHY-1151392. The research of L.S. leading to these results has received
funding from the European Research Council under the European Union's Seventh Framework Programme (FP/2007-2013) / ERC Grant Agreement n. 279972 ``NPFlavour''.
The research of W.A. and S.G. at Perimeter Institute is supported by the Government of Canada through Industry Canada and by the Province of Ontario through the Ministry of Economic Development \& Innovation.  We would like to thank Radovan Dermisek, Jernej Kamenik and Stefan Pokorski for discussions.
We acknowledge support by the Munich Institute for Astro- and Particle Physics (MIAPP) of the DFG cluster of excellence ``Origin and Structure of the Universe''.

\end{document}